\documentclass{egpubl}
\usepackage{EGUKCGVC2024}
 
\WsSubmission      

\usepackage[T1]{fontenc}
\usepackage{dfadobe}

\usepackage{cite}  
\BibtexOrBiblatex
\electronicVersion
\PrintedOrElectronic
\ifpdf 
\usepackage[pdftex]{graphicx} \pdfcompresslevel=9
\else 
\usepackage[dvips]{graphicx} 
\fi
\usepackage[inkscapelatex=false]{svg}

\usepackage{egweblnk} 

\title[Does empirical evidence from healthy aging studies predict a practical difference between visualizations for different age groups?]%
      {Does empirical evidence from healthy aging studies predict a practical difference between visualizations for different age groups?}

\author[Shao et al.]{\parbox{\textwidth}{\centering S. Shao$^{1}$, Y. Li$^{1}$, A.\,I. Meso$^{2}$ and N.S. Holliman$^{1}$}
        \\
{\parbox{\textwidth}{\centering $^1$ Department of Informatics, King's College London, UK\\
         $^2$ Institute of Psychiatry, Psychology \& Neuroscience, King's College London, UK
       }
}
}

\begin{document}
\pdfoutput=1

\maketitle
\begin{abstract}
When communicating critical information to decision-makers, one of the major challenges in visualization is whether the communication is affected by different perceptual or cognitive abilities, one major influencing factor is age. We review both visualization and psychophysics literature to understand where quantitative evidence exists on age differences in visual perception. Using contrast sensitivity data from the literature we show how the differences between visualizations for different age groups can be predicted using a new model of visible frequency range with age. The model assumed that at threshold values some visual data will not be visible to older people (spatial frequency > 2 and contrast <=0.01). We apply this result to a practical visualization and show an example that at higher levels of contrast, the visual signal should be perceivable by all viewers over 20. 
Universally usable visualization should use a contrast of 0.02 or higher and be designed to avoid spatial frequencies greater than eight cycles per degree to accommodate all ages. There remains much research to do on to translate psychophysics results to practical quantitative guidelines for visualization producers.


\end{abstract}  
\section{Introduction}

A significant amount of data is being accumulated in modern society. As a result of the unique capabilities of the human visual system, data visualization enables researchers, analysts, engineers, and general audiences to gain insights into this data more efficiently, as it allows us to detect interesting patterns and features quickly ~\cite{vWij05}.

Among the key issues in visualization is how we ensure visualizations communicate to people when their perceptual or cognitive abilities are different, especially when communicating critical information to decision-makers. One of the important effects is age; there is an increasing body of scientific evidence about how ageing is related to changes that impact our visual abilities. Scientific advances have led to more frequent use of electronic screens in today's progressively ageing society. Considering the universal adoption of electronically displayed visualization, we want to make sure that visualization production is able to account for the visual abilities of all age groups, especially older people. 

An overview of prior work related to ageing and data visualization was provided by ~\cite{WCS24}. This review not only listed the physiological changes in perception, motor control, and cognition that may be affected by ageing and thus the ability of older adults to use visualization effectively but also details specific key points for visual designs. In perception, for example, visual acuity, visual search, colour perception, contrast sensitivity, and hearing with age were described in detail. Some key topics and challenges were investigated. For example, as the complexity of visualizations and tasks increases, older adults' ability to read and comprehend visualizations may be affected by the compounding effects of age-related changes in perceptual, cognitive, and physical abilities. However, it is clear in this broad review that the design of practical and effective data visualizations and interactions for older adults is not supported by enough empirically tested knowledge. In this research, we specifically seek to improve the scientific evidence base for visualizations and to add to evidenced knowledge of what will work in visualization for viewers of different ages starting by considering empirical evidence for contrast sensitivity effects. 

Section 2 provides a brief overview of the literature review on the topics discussed. Section 3 details our research questions. Section 4 discusses the methodology for the topic. Section 5 presents the results of the research problem and the application to data visualization is described in Section 6. In Section 7, challenges and future work are discussed. The conclusion is drawn in Section 8.
\section{Literature Review}
As society ages, it is increasingly important to provide visualization services that are equally accessible for people of all ages given the high dependency on electronic device use today. It is therefore important to investigate the conditions that are needed for creating universally comprehensible visualization and to consider how visualization design can be adjusted if differences in vision are present with age. In Section 2.1, we review the literature on age-related computer visualization designs. A detailed search of the psychological literature in Section 2.2, focusing on contrast sensitivity, for evidence of the relationship between visual perception and age is then conducted.

\subsection{A Review of Age-Related Literature in Computer Science}

Many articles in the field of computer science have assessed older adults' performance when using information visualization. Some work has focused specifically on visual-related capabilities. For example,  ~\cite{BKB07} tested older adults' visual search abilities by using a dynamic display that can present moving dots. Participants were asked to give a response whenever they detected a new dot appearing on the screen. They found that older adults generally performed more slowly and less accurately in detecting the onset of the new dots. They also found that more frequent eye movement during the visual search related to lower accuracy in detection. It was found that in the group of active searchers who made more eye movement during visual search, the age difference in accuracy was larger than in the group of passive searchers who moved their eyes less often.

In addition to visual search, ~\cite{KSvG*02} investigated the age effect in the task that requires both visual ability and kinematic movements. 15 older adults with an average age of 68 years and 15 younger adults with an average age of 23 years were asked to participate in a speed-accuracy task, which was about aiming at targets on the computer screen. Compared to younger adults, older adults performed less well on adapting to the change of target size and moving amplitude of the cursor during the aiming movement. A limitation of the study was that the researchers were only trying to conclude whether older people performed differently from younger people; therefore it did not cover all age groups.

Some studies have explored age-related graphical user interface (GUI) design, examining the differences between the interface designs favoured by older and younger people. Thirty-six older adults and 36 younger adults were asked to view the results of a hypothetical self-monitoring test and perform verbatim comprehension and value interpretation tasks on a computer monitor. The information was represented in basic, colour-enhanced, colour/text-enhanced, and personalized information-enhanced formats. By analyzing the task performance and eye movement data, the researchers found that older adults performed less well on all graphic format tasks as they took longer time to complete and made more errors ~\cite{TYQ18}.

Work reported in ~\cite{vWAD*21} assessed preference and understanding of different types of graphs presenting health risk information by asking people in two age groups to choose the set of graph formats that appealed to them most based on their first impressions. Most participants preferred clock, pie, or bar charts. Both groups understood bar charts fairly well. Neither the younger nor the older participants understood pies very well, and the older participants did not understand clocks like the younger group did.

~\cite{WBG*24} focused on how the type of visualization and number of data points would affect the glanceable perception of people of different ages, especially on a small screen like a smartwatch. The experiment used three charts, bar, donut, and radial, to show three different levels of information, 7, 12, and 24, respectively. The results were analyzed in terms of response time threshold, accuracy, preference and confidence. It was finally concluded that bar and donut are more suitable for quick value comparisons, while radial is more suitable for displaying task progress and completion. Among the possible causes of the gap between different age groups were working memory, stimulus exposure time, and dynamic contexts. It is important to note that the article does not seek to establish underlying performance differences in perception and cognition, it is therefore difficult to conclude which visual adjustments may be most helpful.

The results of these studies point to the fact that ageing may have negative effects on performance while using visualizations, but none of these studies put their focus on the very first stage of performing the task, which is seeing by detecting and encoding the information within the perceptual system during the task. Most researchers assumed that participants could see everything that was supposed to be seen on the screen. However, older people may not be able to see and therefore perceive the necessary information, which could be a possible explanation for the poor performance of older adults in visualization tasks ~\cite{GST06}. In ~\cite{NO03}, participants were tested on their visual search accuracy, time, and number of eye movements under different luminance contrasts. They found that although the performance was resistant to moderate change under high contrast conditions, when the contrast was under a certain level, the deterioration of the performance became more significant. In other words, the inability to perceive some details of the interface can be a significant factor that affects users' performance while using the interface. In addition, questionnaires and feedback methods do not provide a good quantitative result of what age demands of visualization design. We, therefore, move on to look at contrast sensitivity evidence in some detail.

\subsection{ Age-related psychology contrast sensitivity tests}

To explore age-related vision changes and how they can affect visual task performance, we focused on the extensive studies that examined age-related changes in visual function. In 1884, ~\cite{Gal85} measured the eyesight of upwards of 7,000 people, which was influential as it was a large sample size and covered an age range from 6 to 81 years old, participants were asked to look at the writing on a wood block at different distances to test their visual abilities. Although the method was relatively primitive, Galton found a negative relationship between age and eyesight. Unfortunately, it is difficult to convert this data to a usable form for comparison with more modern studies.

After Galton's work, many studies focused on how age affects visual abilities. In ~\cite{Ows16}, Owsley concluded three age-related visual abilities decline: spatial contrast sensitivity, scotopic function, and visual processing speed, which were also the most common causes of performance deterioration. The scotopic function is about the vision under low luminance conditions, the visual processing speed is the speed of taking the visual information to correct judgment, whereas the spatial contrast sensitivity refers to one's ability to perceive the detail of the visual elements under different vision contrast conditions. Compared to scotopic function and visual processing speed, spatial contrast sensitivity is most likely to affect users' capabilities to perceive necessary information, as it represents the ability to detect visual details at different spatial scales. 

Spatial contrast sensitivity refers to the level of contrast a person requires to see a target ~\cite{Ows03}. In this concept, the contrast refers to the perceivable difference in brightness, which reflects the psycho-physical relationship between luminance and brightness perception ~\cite{Nun01}. In the context of the task on a computer monitor, luminance refers to the amount of light emitted by the screen, which is a physical unit (\(\mathrm{cd/m^2}\)) that can be measured directly. Whereas in psychology and vision science, brightness is a subjective perception of the intensity of light ~\cite{Nun01}. This perception does not change linearly with the physical luminance, instead, they show an exponential relationship, which is known as the Stevens' Power Law ~\cite{Ste61}, and also perceived changes relative to surrounding features.

To represent this relationship between the physical feature of light intensity and the subjective perception of brightness, vision scientists and psychologists have devised various ways to calculate such transformation ~\cite{Nag00}. Here, we adopt the Michelson contrast equation ~\cite{Mic95} as shown in Equation \ref{eqn: Michelson}.

\begin{equation}
\label{eqn: Michelson}
Contrast=  \frac{lum_{Max} - lum_{Min}}{lum_{Max} + lum_{Min}}
\end{equation}

This equation defined contrast as the difference between maximum and minimum luminance divided by their sum, where $lum_Max$ is the maximum luminance in a displayed pattern and $lum_Min$ is the minimum luminance in the pattern. By standardizing luminance contrast, this equation provides a measure of visual contrast, which correlates with perceived brightness differences ~\cite{Nag00}. Since Michelson contrast used the peak value to define the luminance contrast, it is particularly effective in describing the periodic pattern such as the Gabor patch ~\cite{KRT*93}. Therefore, this equation is generally adopted by contrast-related psychophysical studies that heavily rely on periodic stimuli ~\cite{Ows16}, ~\cite{SMW*13}. 

Besides psychophysical studies, Michelson's contrast can also be applied in real-life settings. For reference, we tested the Michelson contrast of various types of screens in a typical office environment.  In Table \ref{tab: monitor data}, each screen's maximum luminance (displaying white colour) and minimum luminance (displaying black colour) were tested by a Sekonic L-758Cine light meter using a one-degree spot measurement. The visual contrast was then calculated using Equation \ref{eqn: Michelson}—the calculated contrast representing the maximum Michelson contrast that can be achieved on each monitor in this room. The ambient light in the room (monitor illuminance from other sources) measured with a 180-degree hemisphere was 320 Lux.

\begin{table}[h!]
\centering
\small
\begin{tabular}{|c|p{10mm}|p{10mm}|c|}
\hline
Monitor type & $lum_{Min}$ \newline (\(\mathrm{cd}\)) & $lum_{Max}$ \newline (\(\mathrm{cd}\)) & Michelson contrast \\
\hline
Samsung TV 42" & 9.8 & 270 & 0.93 \\
\hline
Dell laptop 17" & 16 & 550 & 0.94 \\
\hline
Dell 65" & 14 & 260 & 0.90 \\
\hline
Samsung Phone S22  & 20 & 400 & 0.91 \\
\hline
\end{tabular}
\caption{The maximum and minimum luminance of different monitors in a common office setup. The Michelson contrast was calculated using Equation \ref{eqn: Michelson}.}
\label{tab: monitor data}
\end{table}

\begin{figure}
    \centering
    \includegraphics[width=1\linewidth]{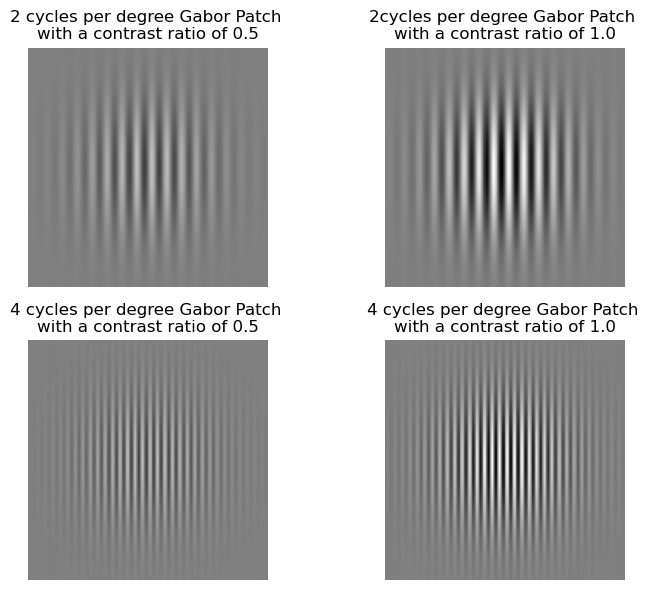}
    \caption{Spatial frequency vs Contrast: examples with Gabor patches.}
    \label{2x2 grid}
    \vspace{-0.8cm}
\end{figure}

When measuring contrast sensitivity, chart-based systems are generally used in the field of vision science and psychophysics. In these charts, sine-wave gratings within a Gaussian window (i.e., the Gabor patch) or letters are often used as test targets. Figure \ref{2x2 grid} shows the example of Gabor patches with different spatial frequencies under different visual contrasts. The density of these periodic gratings in the Gabor patch is called 'spatial frequency', which describes the number of sine-wave cycles divided by visual angle. Since the visual angle of an object depends on viewing distance and the object's size ~\cite{Joy49}, the spatial frequency of visual elements may change according to the display settings. 

Typically, contrast thresholds are expressed on a logarithmic scale of 10 for the contrast level needed for a person to see a target. In research, contrast thresholds are often transformed to contrast sensitivity, which is simply the reciprocal of the threshold. It is referred to as a contrast sensitivity function when contrast sensitivity is calculated as a function of spatial frequency.

There is substantial evidence in the vision science literature showing that older adults' contrast sensitivity (CS) at medium to high spatial frequencies begins to decline at age 30 and progresses into old age ~\cite{SOH82} ~\cite{OSS83}. ~\cite{HSB99} found that spatial visual acuity measurements taken under reduced visual contrast showed a significant decrease in most older adults. Although the experimental population was missing the younger control group, there is still a significant age-related decline in contrast sensitivity within the range from 58 to 102. ~\cite{ZS95} also found that age-related differences in contrast thresholds were significant at low background luminance, but the fact that only 0.5 cycles per degree gratings were used and not tested with a variety of gratings is a shortcoming. 

Experiments by ~\cite{MVO*16} give similar conclusions: as a person ages, contrast sensitivity decreases gradually, with larger decrements at higher spatial frequencies. The level of contrast sensitivity remains almost unchanged until the age of 50, after which it decreases gradually. This research has also shown that low-contrast stimuli are difficult to differentiate. Many older adults underperform when it comes to spatial vision acuity measured under conditions of reduced contrast or luminance, whereas high-contrast visual acuity is inaccurate in assessing the level of functional vision impairment. Furthermore, this study indicated that parameters related to low-contrast stimulation and scattered light change with age, especially when it comes to visual function, optical quality, and intraocular scattering. In the first decades of adulthood, sensory or perceptual factors can compensate for visual deficits, since visual function changes less than objective output, especially before age 50. 

In ~\cite{MVO*16}, the contrast sensitivity was measured using the contrast sensitivity chart CSV-1000E. The test used four rows of sine wave gratings to identify spatial frequencies of 3, 6, 12, and 18 cycles per degree at a distance of eight feet (2.5 meters). However, the CSV-1000 was considered to have some limitations in that it can be unreliable in testing adult and child sensitivity. Using the recommended standardized test protocol, all four reliability estimates of the CSV-1000 chart indicate low reliability for both children and adults. A single examiner's retest record would increase reliability, but not substantially ~\cite{KPK12}. ~\cite{TRM*17}also suggested that trials of searching for target items could be completed using different visual search displays on the screen and that contrast sensitivity and personal visual acuity could be recorded using Optovist vision testing equipment which could be used to determine the relationship between perceptual and attentional parameters and performance in processing quantitative data visualizations.

In the studies mentioned above Owsley et al.~\cite{OSS83} conducted an experiment that best meets the needs of our current topic because it covers a wide range of ages and tests multiple spatial frequencies, and the instrumentation used is conventional. The experiment, which covered a wide range of age groups from 20 to 80, was more representative. Contrast sensitivity for the experiment was measured by an Optronix Vision Tester Model 200, a pre-programmed, microcomputer-controlled television display.
The stimuli were static sinusoidal vertical gratings of the following spatial frequencies: 0.5, 1, 2, 4, 8, and 16 cycles per degree. At a distance of 3 meters, the viewable area corresponds to the area of a rectangle measuring 22.00 cm by 28.82 cm. The average luminance of the screen remained constant at 103 cd/$m^{2}$ and the surround luminance was 2 cd/$m^{2}$.

The experiment carried out by Owsley et al. ~\cite{OSS83} tested contrast sensitivity to obtain contrast thresholds. A suprathreshold contrast of 0.2 is first used to present gratings in 1 group to allow observers to predict what pattern will appear during tracking. The preview was created to minimize spatial frequency uncertainty. The test was initiated after three blank screens, followed by a patch of initial subthreshold contrast, which is then gradually and steadily increased by the computer in a staircase experimental procedure. In the test, contrasts increase by 0.0 to 0.002 at random. A maximum contrast of 0.2 was set, and the whole contrast range took 34 seconds to complete.
In response to the pattern, the observer pressed the button, signaling the computer to reduce contrast when it appeared. As long as the pattern remains visible, the observer is instructed to hold down the button. After the release of the button, the contrast was increased by the computer, and the cycle was repeated. In the process, the contrast threshold is reversed 8 times before it is terminated.
The contrast threshold of each of the six spatial frequencies is derived from the geometric mean of the 8 inversions. The ability to perceive subtle variations in brightness and color may be impaired by reduced contrast sensitivity, while higher contrast sensitivity enables individuals to identify finer details and textures in objects.

\begin{figure}
    \centering
    \includegraphics[width=1\linewidth]{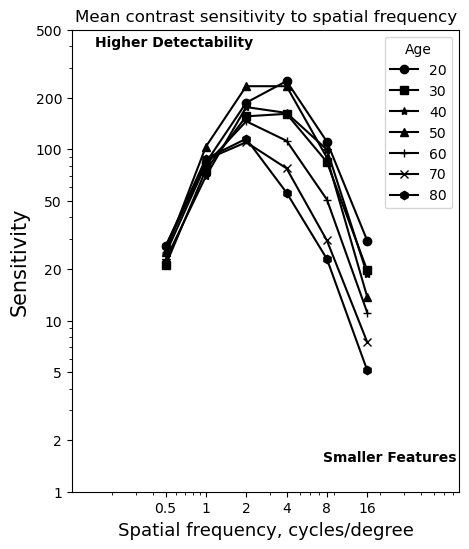}
    \caption{Owsley et al. empirically evaluated the mean contrast sensitivity at each frequency step by age group, figure redrawn from~\cite{OSS83}.}
    \label{fig: Redraw}
\end{figure}

Observing Figure \ref{fig: Redraw}, it appears that the degree of sensitivity to static gratings with low spatial frequencies of less than 1 cycles per degree will remain constant throughout adulthood, whereas the degree of sensitivity to higher spatial frequencies of more than 1 cycles per degree may decrease as we age. Meanwhile, this provides quantitative data on contrast sensitivity and contrast thresholds for different age groups at different spatial frequencies.
\section{Problem Statement}

A key component of perceiving and understanding data visualization is human information processing. When designing and evaluating data visualization, it is essential to consider the normal range of human information processing ability, especially since a number of Western and Eastern societies are rapidly ageing and their reliance on technology is increasing ~\cite{Ver23} ~\cite{VAM18}. When individuals age, their perceptual and motor skills often change, affecting visual ability, which in turn should affect how information and communication technologies are designed. It is important to note that we are considering a normal reduction in visual performance as a result of healthy aging ~\cite{Col19}~\cite{Lib22}. We argue there is a pressing need to reconsider how age influences interaction with computers which means taking empirical evidence of age-related changes into account when designing data visualizations. 

As stated in Section 2, there are a number of general qualitative recommendations concerning GUI design regarding aging, however, very limited empirical quantitative information is available in a form that can guide us. Our research questions are therefore:

\begin{itemize}
    \item What empirical evidence exists in the psychophysics literature that can underpin our understanding of age-related effects?
    \item How can we convert known empirical data to a model that helps us predict when visualizations might not or might not be visible to different age groups?
    \item What do age-related differences look like in practical modern visualizations?
\end{itemize}

For data visualization design to be considered trustworthy across age groups, a better bridge between visualization and psychology must be built, linking to perceptual knowledge and then proceeding to cognitive decision-making.

\section{Method: modelling age-related contrast sensitivity}
Due to the fact that visualization uses an electronic screen to provide information, a number of parameters, such as contrast, brightness, and color, can be adjusted by the user. In addition, the environment a screen is used in will affect the minimum brightness levels due to surface reflections. Contrast provides a way to measure the eyes response to a specific visualization on an specific screen and ultimately the eye's ability to detect the visual information presented on the screen.

As a result of the literature survey in Section 2, we know that contrast sensitivity decreases with age, so contrast and its relation to spatial frequency are of particular importance to characterize.
The data obtained from ~\cite{OSS83} can be analyzed as raw data for modelling the relationship between spatial frequency and age, as illustrated in Table \ref{tab: data} for contrast thresholds at 0.5, 1, 2, 4, 8, and 16 cycles per degree for individuals between the ages of 20 and 80 years old where CT represents contrast threshold and CS means contrast sensitivity. The contrast sensitivity has been converted from these thresholds. Note that the contrast threshold is most often given in the literature in log form. For clarity and in order to restore the original characteristics of the data, we removed the log to obtain the corresponding contrast sensitivity as Table \ref{tab: data} shows.

\begin{table}[]
\footnotesize
\begin{tabular}{|ll|c|c|c|c|c|c|}
\hline
\multicolumn{2}{|l|}{Age/CPD}      & {0.5}   & {1}      & {2}      & {4}      & {8}      & {16}    \\ \hline
\multicolumn{1}{|l|}{20s} & CT & -1.43 & -1.92  & -2.27  & -2.40  & -2.05  & -1.47 \\ \hline
\multicolumn{1}{|l|}{}    & CS & 27.10 & 83.56  & 187.50 & 250.61 & 110.92 & 29.31 \\ \hline
\multicolumn{1}{|l|}{30s} & CT & -1.32 & -1.88  & -2.19  & -2.21  & -1.93  & -1.29 \\ \hline
\multicolumn{1}{|l|}{}    & CS & 20.99 & 76.03  & 155.96 & 161.07 & 84.33  & 19.63 \\ \hline
\multicolumn{1}{|l|}{40s} & CT & -1.35 & -1.84  & -2.25  & -2.21  & -2.00  & -1.29 \\ \hline
\multicolumn{1}{|l|}{}    & CS & 22.59 & 69.66  & 176.60 & 163.31 & 99.77  & 18.58 \\ \hline
\multicolumn{1}{|l|}{50s} & CT & -1.40 & -2.02  & -2.37  & -2.37  & -1.95  & -1.14 \\ \hline
\multicolumn{1}{|l|}{}    & CS & 25.18 & 103.75 & 233.88 & 233.88 & 89.54  & 13.74 \\ \hline
\multicolumn{1}{|l|}{60s} & CT & -1.39 & -1.91  & -2.16  & -2.05  & -1.71  & -1.05 \\ \hline
\multicolumn{1}{|l|}{}    & CS & 24.49 & 81.66  & 145.88 & 112.20 & 50.93  & 11.14 \\ \hline
\multicolumn{1}{|l|}{70s} & CT & -1.42 & -1.94  & -2.04  & -1.89  & -1.47  & -0.87 \\ \hline
\multicolumn{1}{|l|}{}    & CS & 26.30 & 87.10  & 110.15 & 77.45  & 29.38  & 7.48  \\ \hline
\multicolumn{1}{|l|}{80s} & CT & -1.43 & -1.95  & -2.06  & -1.75  & -1.36  & -0.71 \\ \hline
\multicolumn{1}{|l|}{}    & CS & 27.10 & 88.31  & 115.35 & 55.85  & 22.91  & 5.12  \\ \hline
\end{tabular}
\caption{Contrast threshold (CT) of different ages and corresponding contrast sensitivity (CS) derived from ~\cite{OSS83}.}
\label{tab: data}
\end{table}

In our project, we focus on understanding how different viewing abilities on different screens lead to different effects on the characters that should be displayed. In order to calculate actual limits we need to define a specific physical screen and viewing situation, in addition we need to know the frequency and contrast of the visualization when shown on the screen.
In the following the contrast of the displayed visualization is assumed to be constant at 0.01 calculated by Michelson's contrast formula, and the observer is viewing at a fixed distance of 30 centimeters. A 14-inch (1920*1080 pixels) HD laptop display is assumed in our calculations which means that the size of the screen is about 31.0cm * 17.4cm (59.2 by 33.2 degrees of visual angle at 30cm viewing distance).

\begin{figure}
    \centering
    \includegraphics[width=1\linewidth]{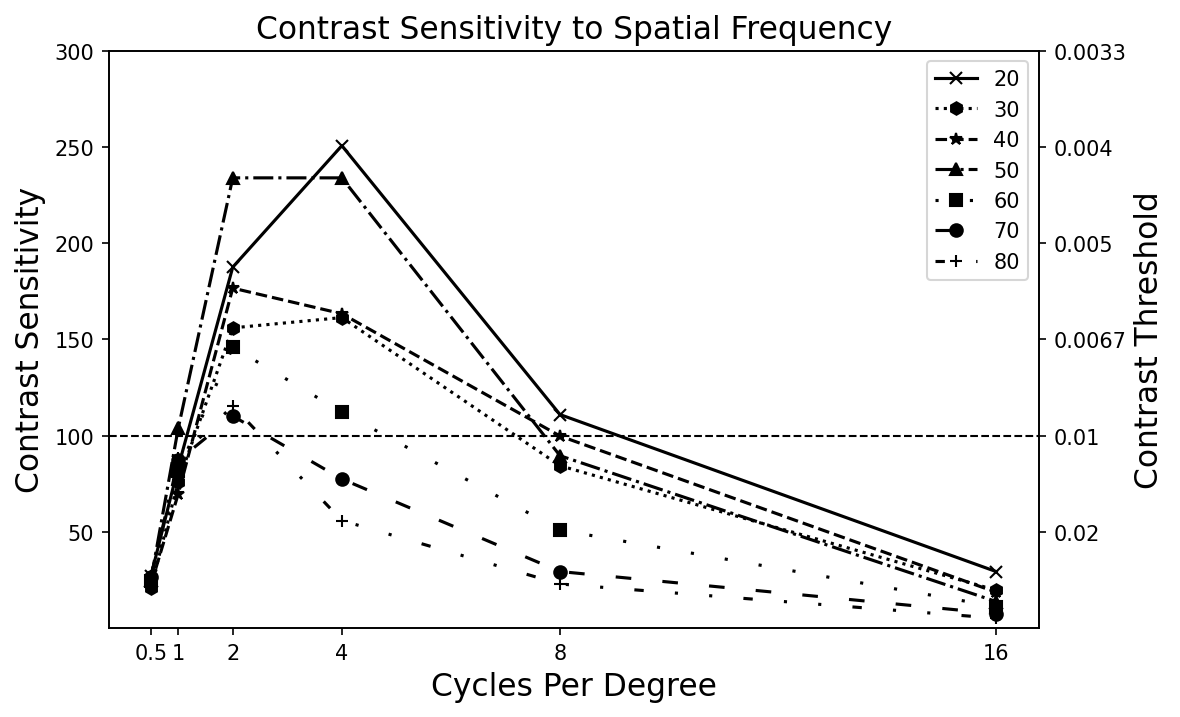}
    \caption{Contrast Sensitivity to Spatial Frequency: any visual signal above the 0.01 line is visible, and the range of signal frequency content visible varies with age group.}
    \label{fig: OurModel}
    \vspace{-0.5cm}
\end{figure}

From Figure \ref{fig: OurModel} we can see that for a visualization contrast of 0.01, each age group has two intersections with the transect line. These two intersections are the lowest visible spatial frequency and the highest visible spatial frequency for each age group. Anything above the transect line is visible for each age group, and we can see that the range of visible frequencies is greater for younger viewers.

A common way to represent a single spatial frequency is the Gabor patch. The Gabor patch is a two-dimensional visual pattern comprised of a sinusoidal wave of specified frequency and orientation modulated by a Gaussian envelope. The bi-normal distribution is used to create a Gaussian envelope for the Gabor patch, which modulates the sine wave and ensures that its amplitude tapers off smoothly towards its edges. With the mathematical formulation of the Gabor patch, these two components are combined to produce a smooth and localized oscillatory pattern as Equation \ref{eqn: GaborFunction} shows.
\begin{equation}
\label{eqn: GaborFunction}
GaborFunction = c*sin(f*(u*x+v*y))\frac{PDF[dist,{x,y}]}{PDF[dist,{x,y}]_{max}}
\end{equation}
\begin{equation}
\label{eqn: AdjustedGabor}
AdjustedGabor = GaborFunction + 0.5
\end{equation}
where $c$ refers to the contrast of the grating, $f$ is the frequency of the sinusoidal grating, and $\frac{PDF[dist,{x,y}]}{PDF[dist,{x,y}]_{max}}$ stands for the scaling Gaussian envelope using the probability density function (PDF) of the binormal distribution to localize the sine wave grating within a specific region and ensure that the envelope correctly scales the intensity values. Due to its mathematical nature, a sine wave will result in a negative value. However, due to the fact that luminance cannot be negative in physics, it is shifted in order to ensure that the overall luminance is not negative as Equation \ref{eqn: AdjustedGabor} shows. In this way, Gabor patches can be produced for a given signal frequency, $f$, and orientation vector, $\{u,v\}$.

We now move on to look at the results of how our model using the Owsley contrast sensitivity data and  Gabor patches combine to allow us to illustrate different thresholds of contrast sensitivity for different age groups.

\section{Results and discussion}

\begin{figure}
    \centering
    \includegraphics[width=1\linewidth]{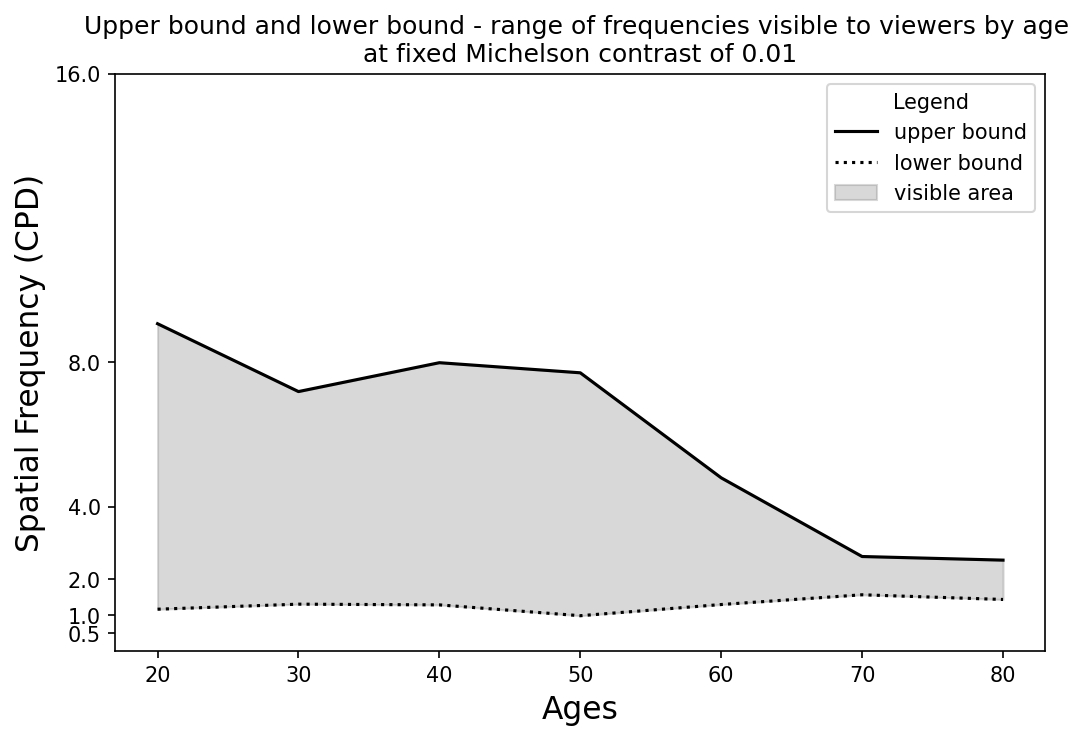}
    \caption{The lower and upper bounds of visible Spatial Frequency with Age at a fixed contrast of 0.01, signals with spatial frequencies within the visible area should be visible. }
    \label{fig: Two bounds}
\end{figure}

As mentioned in the previous section, the horizontal line crosses at two intersections with a set contrast of 0.01, one representing the lowest visible spatial frequency and the other representing the highest visible spatial frequency as shown in Figure \ref{fig: OurModel}. The relationship between all visible low spatial frequencies and age can therefore be described as a lower bound, as shown in Figure \ref{fig: Two bounds}. The same holds true for the upper bound for all high frequencies with age. Based on the diagram, it can be seen that with age the two lines move closer to one another, which also indicates that the middle part of the line is narrowing. The area between the two lines can be considered the 'visible area', which encompasses the range of spatial frequencies that can be seen by individuals of various ages. As anything below or above the range is not perceived by each age group, the region between the two limits becomes the visible spatial frequency area. 

\begin{figure}
    \centering
    \includegraphics[width=1\linewidth]{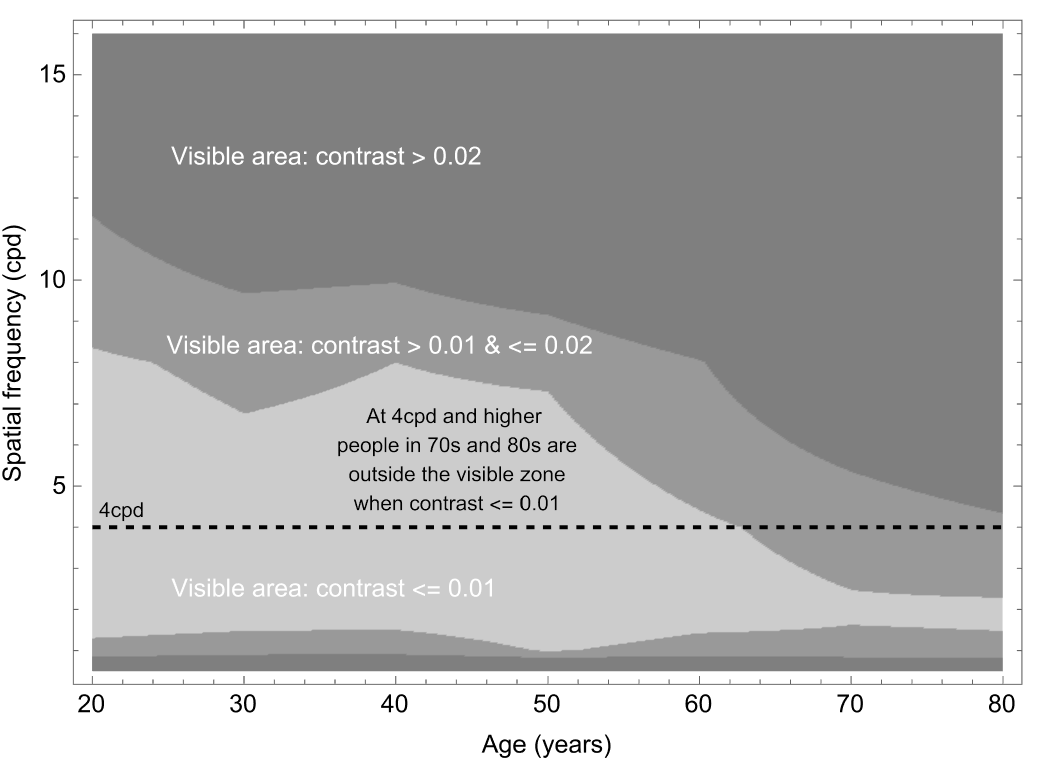}
    \caption{Over the age of 65 and into the 70s and 80s a signal of 4 cycles per degree at a contrast of 0.01 falls out of the visible area and would not normally be discernable by that age group.}
    \label{fig: 70s and 80s}
\end{figure}

We use 2D interpolation of the Owsley data to build a linear model that can predict contrast threshold given a viewer's age and the signal spatial frequency. The model uses Mathematica's default Hermite interpolation with linear (degree 1) interpolating functions, one example of a visualization of its output is illustrated in Figure \ref{fig: 70s and 80s}. This Figure shows how people in their 70s and 80s would be very likely to have difficulty seeing a signal of 4 cycles per degree at a displayed contrast of 0.01. We include the Owsley data and our model in a Mathematica notebook as supplemental material that can be run using the free Wolfram player.

\begin{figure}
    \centering
    \includegraphics[width=1\linewidth]{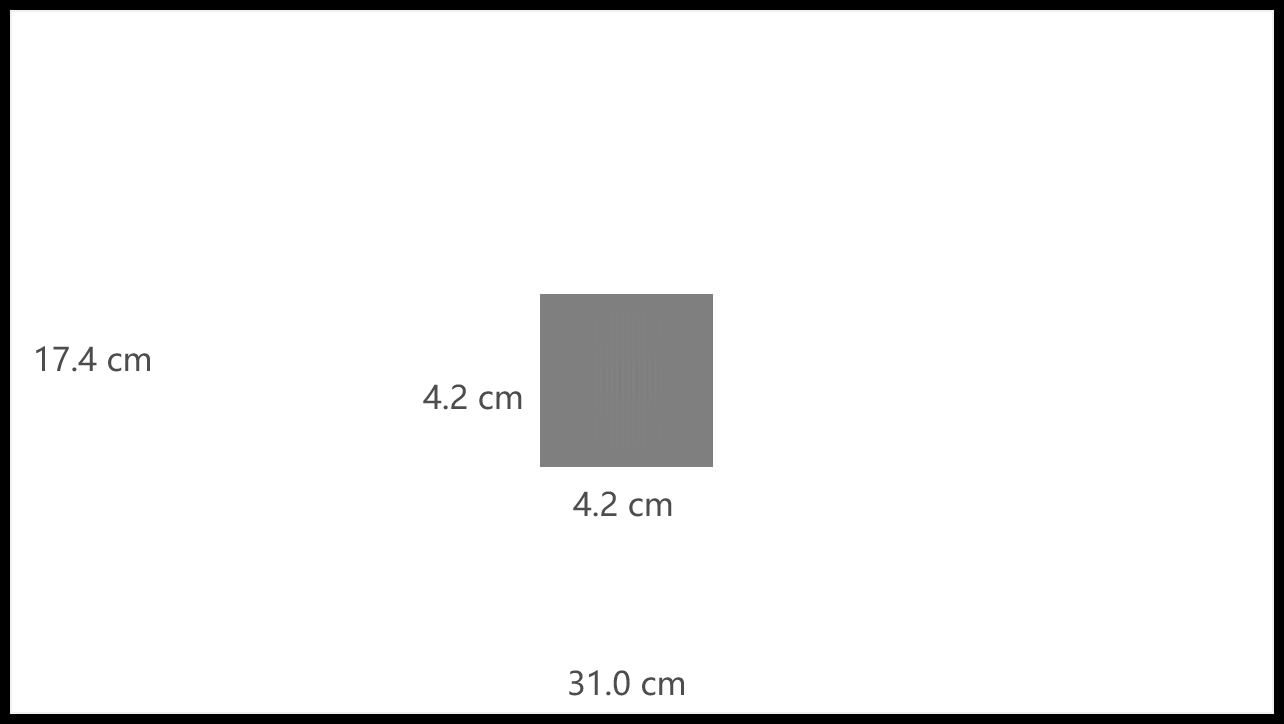}
    \caption{Gabor Patch at 4cpd illustrating the relative size of the patch on the defined screen (not to scale).}
    \label{fig: Gabor Patch on the defined screen}
\end{figure}

There are a variety of factors that can affect the Gabor patch to illustrate glyphs, such as the size of the screen and the viewing distance. Here, we assume the pattern is rendered on the screen of a 14-inch (31.0cm*17.4cm) laptop when viewed at a viewing distance of 300 millimeters from the screen. Figure \ref{fig: Gabor Patch on the defined screen} illustrates the 42mm*42mm Gabor patch for 4 cycles per degree in this scenario. According to our previous analysis, those over 65 and into their 70s and 80 years old will have difficulty distinguishing the signal on the patch, while the rest of the age categories should be able to do so.

It is likely that you realized that the black-and-white Gabor patch in the glyph was not easily discernible to you, which does not suggest that your eyesight is aging beyond your age group. Apart from the lack of consideration given to adjusting the Gabor patch area for your display size (or printout size), image luminance and viewing distance, the purpose of the original psychological experiments was to determine the absolute limits of an individual's ability to see, rather than to simply see clearly. The benefit of this is that it provides a standard for people's perception at different ages and demonstrates we shouldn't show graphics with low contrast of 0.01 and high spatial frequencies like 4 cycles per degree to individuals over 65 and beyond 70 years of age. 

In addition, it shows that there is a spatial frequency limitation for all ages and that too high-frequency graphics, above 8cpd for a contrast of 0.01, should be avoided when designing visualization graphics in order to ensure that they are perceivable by people of all ages.

\begin{figure}
    \centering
    \includegraphics[width=0.5\linewidth]{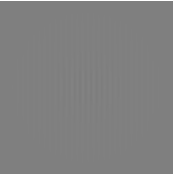}\hfill
    \includegraphics[width=0.5\linewidth]{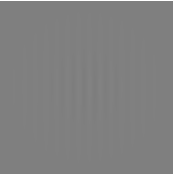}
    \caption{Fixed contrast to 0.01, a 20-year-old can detect a 4 cycles per degree Gabor patch; however, a 70-year-old cannot typically detect it (left). Lower the spatial frequency to 2 cycles per degree, and a 70-year-old will be able to perceive it (right).}
    \label{fig: Gabor0.01}
    \vspace{-0.5cm}
\end{figure}

\begin{figure}
    \centering
    \includegraphics[width=0.5\linewidth]{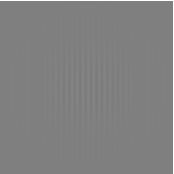}
    \caption{Increasing the contrast to 0.02 will enable all tested ages to perceive 4 cycles per degree.}
    \label{fig: Gabor0.02}
\end{figure}
Using our model based on Owsley data, people of various ages can predict the range of spatial frequencies they can perceive at a given contrast, especially determining whether a Gabor patch of a particular spatial frequency is typically detectable. A 4 cycles per degree Gabor patch with a contrast of 0.01 can be seen by a 20-year-old, whereas a 70-year-old would have difficulty seeing the patch as shown in the left of Figure \ref{fig: Gabor0.01}.  When the spatial frequency drops to 2 cycles per degree (right of Figure \ref{fig: Gabor0.01}), the 70-year-old will be able to see it and the 20-year-old should be able to distinguish it more easily. However, a 4 cycles per degree Gabor patch can be made visible to a 70-year-old by increasing the contrast to 0.02 as shown in Figure \ref{fig: Gabor0.02}, and, the model predicts that this patch is observed at all tested ages.

\section{Application to Data Visualization}

One of the challenges of interpreting psycho-physical experiments in terms relevant to data visualization design is how to apply the empirical results to practical visualizations. In this section, we consider how the results we have from our modeling of age-related visibility from experiments with Gabor patches can be applied to our Vizent glyphs ~\cite{HCF24}.

\begin{figure}
    \centering
    \includegraphics[width=0.25\linewidth]{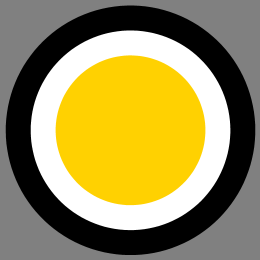}\hfill
    \includegraphics[width=0.25\linewidth]{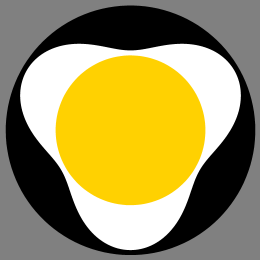}\hfill
    \includegraphics[width=0.25\linewidth]{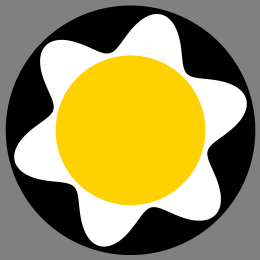}\hfill
    \includegraphics[width=0.25\linewidth]{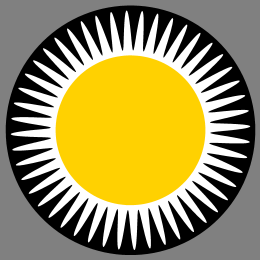}
    \caption{Four Vizent bivariate glyphs~\cite{HCF24}, the central colour represents one variable value, the shape frequency represents a second. They would in real life be displayed as 42mm square patches.}
    \label{fig: Four Vizent bivariate glyphs}
\end{figure}
Figure \ref{fig: Four Vizent bivariate glyphs} showcases the Vizent glyphs, a versatile tool designed to represent bi-variate data. The central colour represents one variable value, while the frequency of the surrounding shape represents a second value. We have successfully applied these glyphs in various data visualization scenarios, including network data visualization ~\cite{AMX*23}, urban sensor visualization ~\cite{HCF24}, and European heatwave data  ~\cite{Bau24}.
\begin{figure}
    \centering
    \includegraphics[width=1.0
\linewidth]{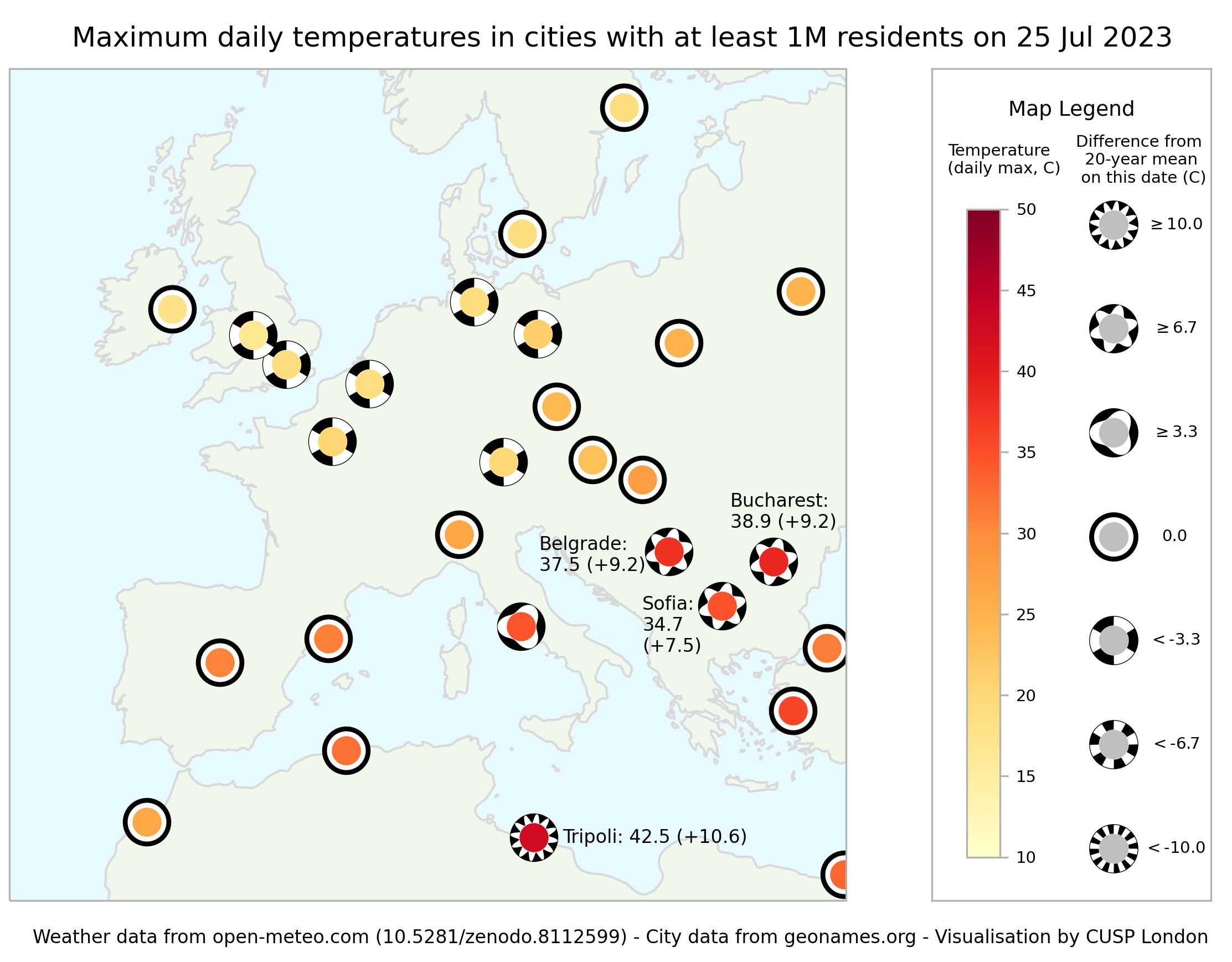}\hfill
  \caption{Vizent bivariate glyphs used to visualize data from the European heatwave data in July 2023. The colour represents daily maximum temperature, the shape frequency how far that temperature exceeds the 20-year mean.}
    \label{fig:EU Heatwave}
\end{figure}

\begin{figure}
    \centering
    \includegraphics[width=0.5\linewidth]{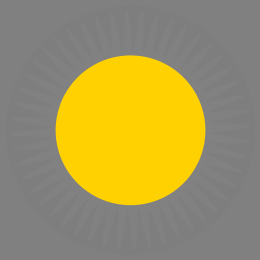}\hfill
    \includegraphics[width=0.5\linewidth]{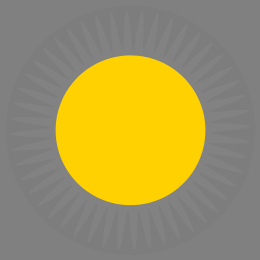}
    \caption{On the left is a glyph which when displayed as a 42mm patch has a dominant spatial frequency of 2.5cpd displayed with 0.01 Michelson contrast, on the right the contrast is raised to 0.02.}
    \label{fig: On the left}
\end{figure}

Considering the fourth shape in Figure~\ref{fig: Four Vizent bivariate glyphs}, generated using a 48 Hz radial sine wave. To find its image characteristics in cycles per degree we use an image space 2D Fourier transform and take the frequency of maximum power as the dominant part of the signal. In this case, this is calculated as 20Hz which is equivalent to 2.5 cycles per degree if displayed on the display described earlier.
Using our model of Figure \ref{fig: OurModel} we can predict this is unlikely to be visible to those over 70 if it was shown at a Michelson Contrast of 0.01. However, they should be able to (just) see it if the contrast in increased to 0.02 or higher. This is illustrated in Figure \ref{fig: On the left}. The highest frequency shape we have analysed is equivalent to just over 8cpd on the target display, and this should not be visible to viewers of any age at a contrast of 0.01.

A similar calculation for the glyphs in the EU Heatwave visualization Figure~\ref{fig:EU Heatwave} shows that when they are displayed at 10mm (which happens when the image is shown on our target display) then the frequency of the most detailed glyph is just under 4cpd. At a low contrast of 0.01 this would not be visible to viewers over 65.

 In practice, we use much higher contrast values than threshold, and displays that are bright enough in office environments to show them, but note there is no routine control over the spatial frequency and contrast of visualization content.

\section{Contributions and Future Work}

Reviewing our research questions we summarise the contributions of this article below: 

Our first question was: \emph{ what empirical evidence exists in the psychophysics literature that can underpin our understanding of age-related effects?}

We have demonstrated that there is relatively little quantitative evidence on ageing in use in visualization but instead widespread use of qualitative and informal guidelines. However, there is a significant and useful set of quantitative empirical studies that we can build on in existing psychophysics literature.

Our second, \emph{how can we convert known empirical data to a model that helps us predict when visualizations might not be or might be visible to different age groups?}

We presented a new interpolating model that re-interprets Owsley's psychophysical results such that we can predict visibility of a visual signal given the signals spatial frequency in cycles per degree and the age of the viewer. We demonstrated how this applies to Gabor patches, typical of those used in psychophysical experiments.

Finally, we considered, \emph{ What do age-related differences look like in practical modern visualizations?} 

Based on our recent work with the Vizent bivariate glyphs we demonstrated how the model limits on spatial frequency and contrast are likely to affect data visualization methods. Since we tend to use high contrast values, compared to threshold limits, our current visualizations should not be affected by age differences on the target display.

One issue that we have not considered is the upper-frequency limit on visual perception and this will also become an issue when visualizations are displayed on smaller screens (such as mobile screens), or larger screens set further away. It is also the case that the contrast and pixel resolution specifications of screens all affect the reproducibility of visualizations and there is a need to review the working range of spatial frequency contrast on those currently in use from projection screens down to mobile screens in future work.

Our current model and conclusions are based on a very strict assumption of a specific screen size and viewing distance to ensure that age is the only variable. We do this so as to be able to assume a display calibration and control of displayed image size when analysing the glyphs. This display calibration is not commonly implemented across practical display devices, and contrast will further vary in different ambient lighting situations. We suggest there is a need empirical work to evaluate definitely noticeable limits, rather than threshold limits, as well as differences of contrast by display and viewing situation. 

The present results pave the way for future work by demonstrating that people's visual ability changes with age, and that visualizations should be designed with spatial frequency in mind to ensure that people of different ages see the intended visual information.

\section{Conclusion}

In conclusion, this paper reviews the literature on the relationship between healthy ageing and vision in the areas of visualization and psychology with the goal of finding ways to connect them to understand whether age-related visualization design guidelines are needed.
Using data from empirical studies we built a predictive model that encodes the range of visual ability for different age groups, our new model clearly illustrates the practical differences between age groups.


\clearpage

\bibliographystyle{eg-alpha-doi} 
\bibliography{CGVC-aging-bib-1}       



\end{document}